\PassOptionsToPackage{dvipsnames}{xcolor}
\documentclass[10pt,sigconf,nonacm,screen]{acmart}

\usepackage[all]{nowidow}
\usepackage{xcolor}
\usepackage{subcaption}
\usepackage{hyperref}
\usepackage{acronym}

\usepackage[capitalise,noabbrev]{cleveref}
\crefformat{section}{\S#2#1#3} 
\crefformat{subsection}{\S#2#1#3}
\crefformat{subsubsection}{\S#2#1#3}

\newcommand{\sysname}{\textsf{Lotus}}

\begin{document}

\author{Minghe Wang}
\affiliation{%
    \institution{TU Berlin \& ECDF}
    \city{Berlin}
    \country{Germany}}
\email{mw@mcc.tu-berlin.de}

\author{Trever Schirmer}
\affiliation{%
    \institution{TU Berlin \& ECDF}
    \city{Berlin}
    \country{Germany}}
\email{ts@mcc.tu-berlin.de}

\author{Tobias Pfandzelter}
\affiliation{%
    \institution{TU Berlin \& ECDF}
    \city{Berlin}
    \country{Germany}}
\email{tp@mcc.tu-berlin.de}

\author{David Bermbach}
\affiliation{%
    \institution{TU Berlin \& ECDF}
    \city{Berlin}
    \country{Germany}}
\email{db@mcc.tu-berlin.de}

\title{\sysname{}: Serverless In-Transit Data Processing for Edge-based Pub/Sub}

\begin{abstract}
    Publish-subscribe systems are a popular approach for edge-based IoT use cases:
    Heterogeneous, constrained edge devices can be integrated easily, with message routing logic offloaded to edge message brokers.
    Message processing, however, is still done on constrained edge devices.
    Complex content-based filtering, the transformation between data representations, or message extraction place a considerable load on these systems, and resulting superfluous message transfers strain the network.
    
    In this paper, we propose \sysname{}, adding in-transit data processing to an edge publish-subscribe middleware in order to offload basic message processing from edge devices to brokers.
    Specifically, we leverage the Function-as-a-Service paradigm, which offers support for efficient multi-tenancy, scale-to-zero, and real-time processing.
    With a proof-of-concept prototype of \sysname{}, we validate its feasibility and demonstrate how it can be used to offload sensor data transformation to the publish-subscribe messaging middleware.
\end{abstract}

\maketitle

\section{Introduction}
\label{sec:introduction}

By bringing cloud-like compute resources closer to users and devices, to the \emph{edge} of the network, edge computing facilitates novel application areas such as the Internet of Things (IoT)~\cite{paper_pfandzelter2021_zero2fog,paper_bermbach2017_fog_vision}.
Combining edge and cloud resources offers applications low-latency data access~\cite{poster_hasenburg2020_towards_fbase}, limits network strain~\cite{paper_bermbach2017_fog_vision}, and ensures privacy compared to cloud-only computing~\cite{paper_pallas2020_fog4privacy,paper_grambow2018_video_surveillance}.

Edge and IoT applications often rely on edge publish-subscribe (pub/sub) middleware which offers a scalable and convenient solution for edge-to-edge communication~\cite{rausch2018emma,happ2020impact,paper_hasenburg2020_geobroker,paper_hasenburg2020_disgb,paper_hasenburg2020_broadcast_groups}.
Here, message routing and delivery logic are handled not on IoT devices but rather on pub/sub brokers, increasing ease-of-use for developers and reducing computational intensity on devices~\cite{paper_hasenburg2020_geobroker}.

While there are some research prototypes supporting content-based filtering beyond topics, e.g.,~\cite{zhang2017efficient, qian2014h, shi2022hem, qian2019adjusting,li2022prid, qian2014rein}, today's pub/sub message brokers essentially treat all messages as black boxes, leaving all data processing to the subscribers.
Such processing can include additional filtering and arbitrarily complex data transformation, i.e., we currently deliver \emph{more} messages than needed~\cite{paper_hasenburg2020_geobroker} and then use \emph{constrained} devices to process them.

Consider the example of a smart home in which smart blinds subscribe to data from an outdoor temperature sensor:
When the blinds receive a sensor reading, they query data from other sensors (e.g., brightness, indoor temperature) as well as user preferences (e.g., temperature thresholds, time) to decide whether to trigger an action or not.
In this example, sensor readings are delivered and processed even if not needed (e.g., no significant change in temperature), resulting in unnecessary network consumption and compute cost.
With \emph{multiple} smart blinds, the \emph{exact same} superfluous processing is even done several times in parallel.

In this paper, we propose \sysname{}, a pub/sub middleware for the edge that integrates in-transit data processing.
This way, data is still processed in real-time, on the shortest path from sender to recipient~\cite{paper_pfandzelter2019_functions_vs_streams} but (i) only the messages actually needed by the subscriber are delivered, (ii) messages are delivered in the format that is needed by the subscriber, and (iii) data processing is run only once per message.
Essentially, we allow edge devices to subscribe to $f(x)$ rather than the topic $x$ and process events for multiple subscribers only once.
To support arbitrarily complex filtering and data transformation, we leverage the fact that Function-as-a-Service (FaaS) can process messages in real-time, in an isolated manner, and with efficient scalability on the edge broker~\cite{gadepalli2019challenges,hall2019execution,paper_bermbach2020_auctions4function_placement,paper_bermbach2021_auctionwhisk,paper_pfandzelter2020_tinyfaas}.
\sysname{} increases ease-of-use for developers, because they will receive only those data they need and in the format they need.
At the same time \sysname{} also increases resource efficiency as it delivers less unneeded messages content with data extraction.

To this end, we make the following contributions:

\begin{itemize}
    \item We describe the design of \sysname{}, an edge pub/sub communication middleware with support for in-transit data processing (\cref{sec:systemdesign}).
    \item We present a proof-of-concept prototype (\cref{sec:implementation})\footnote{We make our prototype implementation available as open-source software: \url{https://github.com/Mhwwww/Lotus}.}.
    \item We demonstrate the feasibility of \sysname{} in three use-cases (\cref{sec:feasibility}).
\end{itemize}

\section{Background \& Related Work}
\label{sec:background}

We start with an outline of the concepts of pub/sub systems (\cref{sec:background:pubsub}) and FaaS platforms (\cref{sec:background:faas}), with a focus on their application in edge computing.
We then give an overview of related work (\cref{sec:background:relwork}).

\subsection{Edge Communication with Pub/Sub}
\label{sec:background:pubsub}

Pub/sub is an n-to-m messaging pattern.
In practice, pub/sub is usually broker-based and topic-based.
This means that (i) participants communicate through brokers rather than exchanging messages directly~\cite{tarkoma2012publish} and
(ii) \emph{publishers} publish messages to a \emph{topic} that \emph{subscribers} subscribe to, receiving the topic's messages.
Note that any physical device can act as both publisher and subscriber to multiple topics.

Pub/sub is commonly employed in edge computing as it provides scalable, low-latency, resilient, and scalable communication between heterogeneous edge devices~\cite{rausch2018emma,happ2020impact,paper_hasenburg2020_broadcast_groups}.
In this geo-distributed environment, low communication latency and high scalability can be achieved through distributing the broker and limiting message dissemination with efficient routing.
\emph{GeoBroker}~\cite{paper_hasenburg2020_geobroker,paper_hasenburg2020_disgb} uses geographic context information on publishers and subscriptions in order to route messages only to areas of interest.
Using distributed \emph{rendezvous points} close to clients, messages are processed with low latency.

\subsection{FaaS at the Edge}
\label{sec:background:faas}

FaaS is a cloud computing model in which cloud providers manage and run individual functions in response to specific events or requests~\cite{baldini2017serverless, mcgrath2017serverless}.
The abstractions of FaaS are beneficial to edge computing, as its fine-grained, on-demand resource allocation supports a more efficient use of the limited edge resources.
Further, the higher level of abstraction for developers also helps abstract from challenges of edge computing such as geo-distribution, as the responsibility for managing edge characteristics is shifted to the edge FaaS platform~\cite{gadepalli2019challenges,hall2019execution,paper_bermbach2020_auctions4function_placement,paper_bermbach2021_auctionwhisk,paper_pfandzelter2021_LEO_serverless}.

A number of edge-focused FaaS platforms have been proposed, including abstractions for microcontrollers~\cite{george2020nanolambda} and the entire edge-fog-cloud continuum~\cite{paper_bermbach2021_auctionwhisk}.
\emph{tinyFaaS}~\cite{paper_pfandzelter2020_tinyfaas} is a FaaS platform for small- and medium-sized single node systems designed as a building block for larger platforms.
By removing the components needed to build cloud and hyperscale FaaS platforms such as OpenWhisk~\cite{breitgand2018lean} and their associated overhead, tinyFaaS can achieve a small resource footprint and improved performance.

\subsection{Related Work}
\label{sec:background:relwork}

There exist some preliminary approaches to integrate data processing with edge pub/sub systems.
Čilić et al.~\cite{10.1007/978-3-031-20936-9_3} propose an adaptive data-driven routing architecture based on content-based pub/sub to ensure continuous data delivery from IoT devices in the edge-to-cloud continuum.
Arruda et al.~\cite{10.1007/978-3-030-70866-5_13} use Reinforcement Learning based on a topic-based pub/sub system to achieve efficient data distribution, especially for the restricted edge environment.
Huang et al.~\cite{HUANG2020167} propose a multi-tenant blockchain enhanced communication model based on pub/sub to achieve system security at the edge by exploiting the salient features of blockchain.
Li et al.~\cite{li2018fog} focus on structured pub/sub for Internet of Vehicles, using boolean expressions to process data.
They further consider the spatial requirements of fog environments to enable efficient indexing and matching.
These approaches for data processing for pub/sub at the edge improve performance, but scalability, multi-tenancy, and resource efficiency constraints are not a focus.

Incorporating the FaaS paradigm could solve these issues, but the combination of pub/sub and FaaS has so far only been considered in a cloud context.
Nasirifard et al.~integrate pub/sub systems in IBM Bluemix, AWS Lambda and OpenWhisk~\cite{hafeez2018demo, nasirifard2017serverless, nasirifard2022serverless}.
Their proposed systems perform topic-based, content-based and function-based matching based on the FaaS paradigm.
This function-based matching takes user-defined functions and applies them to cloud publications, only the publications which can pass the function logic will be forwarded to the subscribers.
This work shows the feasibility of applying FaaS to a pub/sub system in the cloud and demonstrates the scalability that FaaS could bring.
The proposed function-based matching, however, leaves out the popular topic-based matching and uses function logic only for content-based matching.

\section{\sysname{} Architecture}
\label{sec:systemdesign}

While the message routing logic can be offloaded to an edge broker in pub/sub, processing of incoming messages must still be performed on the constrained edge devices.
By moving more of this processing to the edge broker, enabling in-transit data processing in an edge pub/sub middleware, resource use on constrained devices can be decreased.
With \sysname{}, we incorporate FaaS concepts to enable efficient, scale-to-zero, and real-time processing.

We design \sysname{} for four main objectives: (i) Offloading data processing from edge devices to the broker to reduce resource consumption, since the edge devices are usually resource constrained, (ii) Allowing edge devices to subscribe to $f(x)$ rather than the topic $x$ to use the FaaS paradigm for system scalability, flexibility, and variability, (iii) Enabling data extraction to achieve accurate and non-redundant messaging to improve resource efficiency and reduce network strain, and (iv) Leveraging the FaaS paradigm to simplify development.

Based on these objectives, we present the architecture for \sysname{}.
We want \sysname{} on the edge rather than in the cloud because the edge is closer to IoT applications to provide lower latency and less bandwidth-consuming communication.
Since edge devices are located on the outermost layer of a network and are usually geo-distributed, taking geo-context into account results in more accurate data processing.
As the FaaS paradigm allows for the management and execution of isolated functions in response to specific events or requests, it brings scalability and resilience to the system and also supports privacy as data is processed on the edge which is closer to data sources.

\begin{figure}
    \centering
    \includegraphics[width=0.9\linewidth]{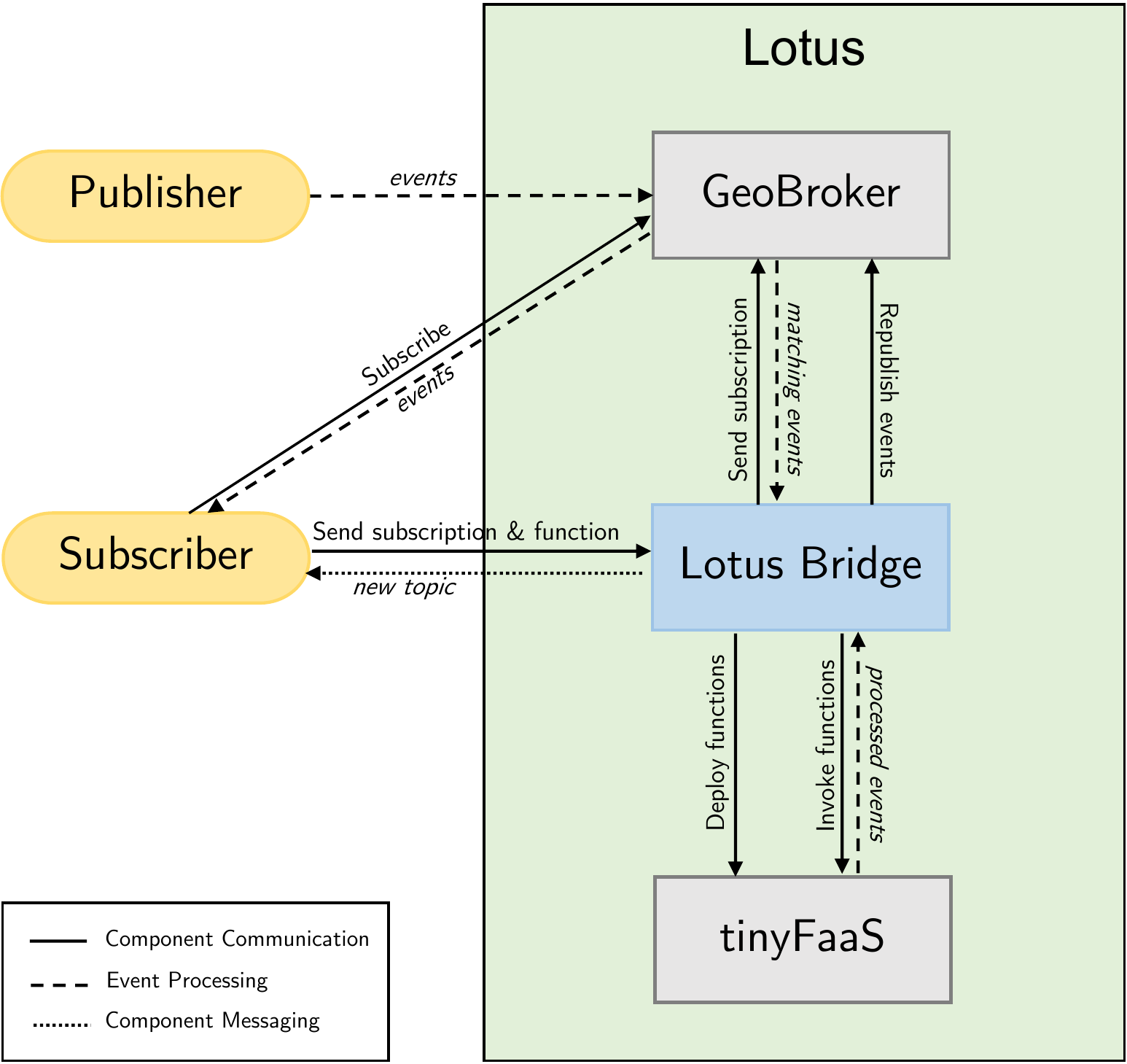}
    \caption{Conceptual Architecture:
        \sysname{} is an edge pub/sub middleware that integrates GeoBroker and tinyFaaS with \sysname{} Bridge to add in-transit data processing with client-specific functions.
        It receives incoming requests, performs geo-filtered event matching in GeoBroker, invokes functions and gets processed results from tinyFaaS.
        \sysname{} Bridge republishes the processed results to GeoBroker which ensures that if multiple subscribers are interested in the same original event set, the events are processed only once.
    }
    \label{fig:systemdesign}

\end{figure}

We show the conceptual architecture of \sysname{} in \Cref{fig:systemdesign}.
We extend the GeoBroker edge pub/sub middleware with support for invoking functions on tinyFaaS for incoming messages.
Subscribers subscribe to topics through the \sysname{} middleware, specifying function code that is executed for every incoming message.
\sysname{} provides three functionalities here, (i) Providing an entry point for the IoT applications to hand over subscriptions and functions, (ii) Adding a processed subscription to GeoBroker,
and (iii) Deploying or removing client-specific functions to tinyFaaS, sending the matching events to function, and forwarding the processed results.
Thus, \sysname{} extends GeoBroker by processing the content of messages in a geo-distributed context, and tinyFaaS allows users to upload arbitrary code and run it in an isolated manner.
As all components are co-deployed on the same physical node, no additional communication latency is incurred.
\sysname{} Bridge republishes the function-processed events to the new topic to which the client is now subscribed, so subscribers receive processed events directly from GeoBroker.
If multiple subscribers subscribe to the same topic with the same processing function, the original event is only processed once.

\section{Demonstration}
\label{sec:demonstration}
To demonstrate the feasibility of \sysname{}, we present a proof-of-concept prototype (\cref{sec:implementation}) and show its application in three use cases: content-based filtering (\cref{sec:feasibility:contentbased}), message transformation (\cref{sec:feasibility:transformation}), and data extraction (\cref{sec:feasibility:extraction}).

\subsection{Proof-of-Concept Implementation}
\label{sec:implementation}

As a proof-of-concept, we implement a prototype of \sysname{} that supports targeted data distribution while considering both the geographical context of clients and user-defined functionality in terms of processing functions.
This means that subscribers will receive processed data only and only in those cases where both geo-context checks of GeoBroker~\cite{paper_hasenburg2020_geobroker,paper_hasenburg2020_disgb} are passed.

\begin{figure}
    \centering
    \includegraphics[width=\linewidth]{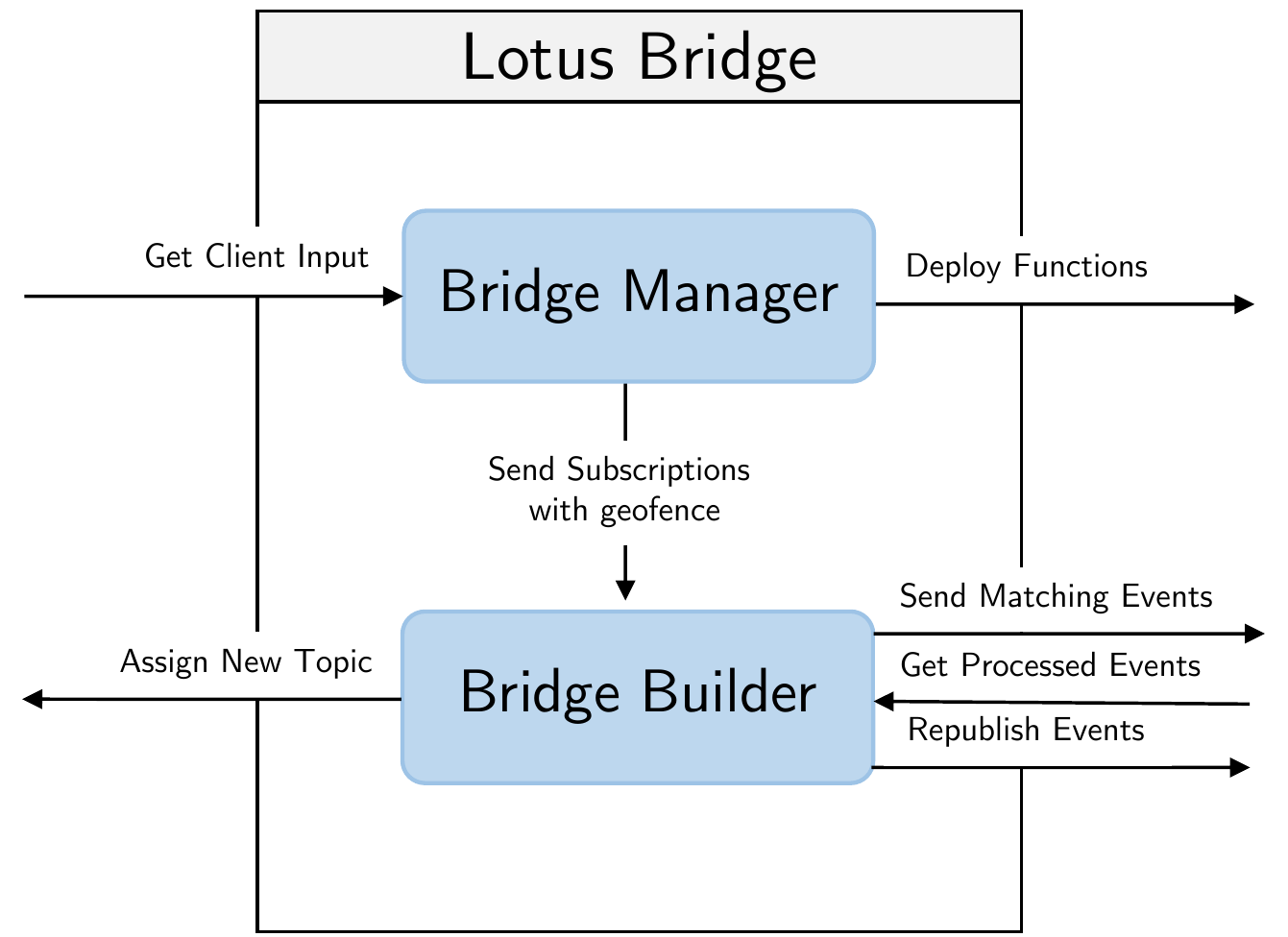}
    \caption{\sysname{} Functionalities:
        The Bridge Manager is responsible for receiving incoming requests and assigning them to GeoBroker and tinyFaaS respectively.
        The Bridge Builder handles event interactions between GeoBroker and tinyFaaS.
    }
    \label{fig:functionalities}
\end{figure}

We show the main functionalities of \sysname{} Bridge in \Cref{fig:functionalities}, the main components are the \emph{Bridge Builder} and \emph{Bridge Manager}:
The Bridge Manager provides an entry point for client requests, while the Bridge Builder is responsible for setting up the connection between GeoBroker and tinyFaaS.
Both of them were implemented in Kotlin.

\subsubsection{Bridge Manager}
The Bridge Manager receives client input, which includes subscription and client-specific functions, and then sends subscriptions to GeoBroker and deploys the functions to tinyFaaS.
Subscriptions and geo-fences are used to invoke the Bridge Builder to subscribe to matching events from GeoBroker.
Clients specify a function deployment package that the Bridge Builder uploads to a local tinyFaaS instance.
Other operations, such as deleting functions and getting function lists, are also supported in the same way.
Afterwards, the client will subscribe to a new topic assigned by the Bridge Builder to wait for the function-processed events.

\subsubsection{Bridge Builder}
To support the maintenance and efficiency of the prototype, we have the Bridge Builder to establish communications between GeoBroker and tinyFaaS. 
The Bridge Builder connects to GeoBroker using the GeoBroker API, forwarding client subscriptions.
When a message is received on a subscription, the Bridge Builder formats the events and sends them to the client function in tinyFaaS over an HTTP request.
If a processed event is returned, it will be republished on GeoBroker with a new topic that the clients have subscribed to, and clients could get the desired events directly from GeoBroker.
In this way, if several devices with the same function satisfy the geo-constraints, which means they are interested in the same original events, they can subscribe to the same new topic to get targeted events.
Thus, the original events are only processed once for multiple subscribers.

\subsection{Use Case Scenarios}
\label{sec:feasibility}

To demonstrate how \sysname{} can be used, we implement several scenarios using our prototype.
All experiments are built in Kotlin, only deployed tinyFaaS functions are written in Node.js.
We perform all experiments on a MacBook Pro with an M1 processor.

\begin{figure*}
    \centering
    \includegraphics[width=0.9\linewidth]{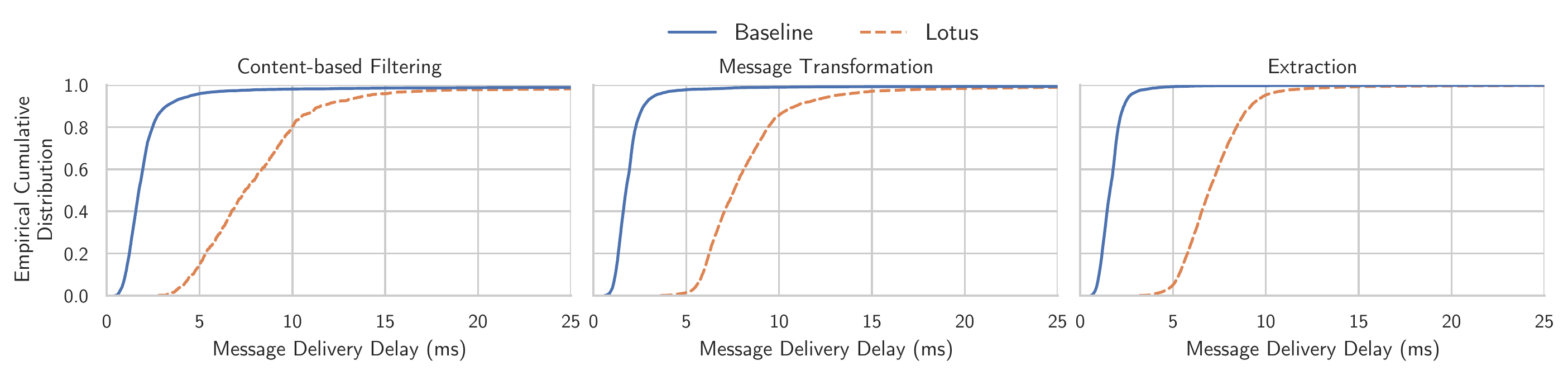}
    \caption{The additional processing overhead incurred by message processing on tinyFaaS in \sysname{} is in the single-digit millisecond range. Compared to possible savings in bandwidth and edge device resources, we consider this a viable trade-off.}
    \label{fig:demonstration}
\end{figure*}

\subsubsection*{Content-based Filtering}
\label{sec:feasibility:contentbased}

One use case for \sysname{} is filtering messages not only based on their topic and location, but also by applying rules to their content to determine which subscribers they should be forwarded to.
Subscribers can upload a function that matches messages with a user-specified rule set, e.g., a specific field needs to exceed a threshold.
This way, edge network resources and computing power on resource constrained edge devices can be saved by only delivering relevant messages.

To showcase this, we have implemented a scenario in which edge sensors continuously produce weather measurements, e.g., temperature and wind speed.
Other edge devices are subscribed to these measurements to show warnings to users in case of extreme weather.
Since most weather measurements are not extreme, edge devices immediately discard almost all incoming messages without doing anything with them.
This puts unnecessary load on the network and uses sparse resources on these edge devices.
With \sysname{}, subscribers can upload a function that filters all incoming weather measurements for extreme weather, and only forwards these.
We implement this use case with 50 publishers (100 published messages each) and 50 subscribers, which filter out 79\% of messages.
This greatly reduces the load on the network and edge devices.
As shown in \cref{fig:demonstration}, with invoking the filtering function on \sysname{}, average message delivery time is increased from 2.54ms to 8.59ms.

\subsubsection*{Message Transformation}
\label{sec:feasibility:transformation}

Another use case is transforming messages from one data format into another by applying a transformation function.
This can be useful if producers or consumers are very resource-constrained, or if their software is not under the control of the users.
In our showcase, producers produce a JSON list, while subscribers can only work with lists that are in CSV format.
To translate between the two systems, a function can be uploaded to \sysname{} which takes the JSON as input and returns a CSV representation.
We implement this use case with 50 publishers (100 published messages each) and 50 subscribers.
Depending on the complexity of the data, this only adds a small amount of additional processing delay (average 6.39ms) to every event, as shown in \cref{fig:demonstration}.

\subsubsection*{Data Extraction}
\label{sec:feasibility:extraction}

In some cases, subscribers might only be interested in a very small part of the whole message that is published.
\sysname{} allows subscribers to limit the amount of data sent with every event by applying data extraction: Instead of forwarding the whole message, only the relevant parts are transmitted, decreasing network usage and computational overhead on the receiving side.
To showcase this, we have implemented a use case where 50 publishers publish 100 JSON objects with one key that is important for 50 subscribers, and 100 additional keys that are not relevant for the subscribers.
By uploading a function that condenses the message down to the important parts, subscribers can reduce their message size (in our case to just 1\% of the original message size).
As shown in \cref{fig:demonstration}, the message delivery delay increases from an average 1.74ms in the baseline to 7.26ms with \sysname{}.

\section{Conclusion \& Future Work}
\label{sec:conclusion}

We have presented \sysname{}, an edge publish/subscribe middleware that adds in-transit processing and leverages the FaaS diagram to offload processing from edge devices to brokers in order to reduce network strain and optimize resource utilization.
We implemented a proof-of-concept prototype and demonstrated it in three use cases: content-based filtering, message transformation, and data extraction.

Here, we have shown the feasibility of incorporating the FaaS paradigm into pub/sub systems in a single node environment.
In future work, we plan to extend the approach and prototype for distributed deployments, basing it on DisGB rather than the single-node GeoBroker.

\begin{acks}
    Funded by the \grantsponsor{BMDV}{Bundesministerium für Digitales und Verkehr (BMDV, German Federal Ministry for Digital and Transport)}{https://bmdv.bund.de/EN/} -- \grantnum{BMDV}{19F1119A}.
\end{acks}

\balance
\bibliographystyle{ACM-Reference-Format}
\bibliography{bibliography.bib}

\end{document}